\documentclass{article}
\usepackage{spconf,amsmath,graphicx}

\usepackage{enumitem}
\usepackage{amsfonts}
\usepackage{booktabs}
\usepackage{hyperref}
\usepackage{array}
\setlist{nosep, leftmargin=14pt}

\def\eg{{\emph{e.g.}}}
\def\ie{{\emph{i.e.}}}

\title{Differentiable Projection from Optical Coherence Tomography B-Scan without Retinal Layer Segmentation Supervision}
\name{Dingyi Rong$^{1,\star}$\thanks{$^{\star}$Dingyi Rong and Jiancheng Yang have contributed equally.} \qquad Jiancheng Yang$^{1,2,\star}$  \qquad Bilian Ke$^{3,\dagger}$\thanks{$^{\dagger}$Corresponding author: Bilian Ke.} \qquad Bingbing Ni$^{1}$}

\address{$^{1}$ Shanghai Jiao Tong University, Shanghai, China \\
$^{2}$ EPFL, Switzerland\\
$^{3}$ Shanghai General Hospital, Shanghai, China\\
{\tt\small \{r892546826, jekyll4168, nibingbing\}@sjtu.edu.cn, kebilian@126.com}
}

\begin{document}

%
\maketitle
\begin{abstract}
Projection map (PM) from optical coherence tomography (OCT) B-scan is an important tool to diagnose retinal diseases, which typically requires retinal layer segmentation. In this study, we present a novel end-to-end framework to predict PMs from B-scans. Instead of segmenting retinal layers explicitly, we represent them implicitly as predicted coordinates. By pixel interpolation on uniformly sampled coordinates between retinal layers, the corresponding PMs could be easily obtained with pooling. Notably, all the operators are differentiable; therefore, this Differentiable Projection Module (DPM) enables end-to-end training with the ground truth of PMs rather than retinal layer segmentation. Our framework produces high-quality PMs, significantly outperforming baselines, including a vanilla CNN without DPM and an optimization-based DPM without a deep prior. Furthermore, the proposed DPM, as a novel neural representation of areas/volumes between curves/surfaces, could be of independent interest for geometric deep learning. 
\end{abstract}
\section{Introduction}
\label{sec:intro}

\begin{figure}

\begin{minipage}[b]{1.0\linewidth}
  \centering
  \centerline{\includegraphics[width=8.5cm]{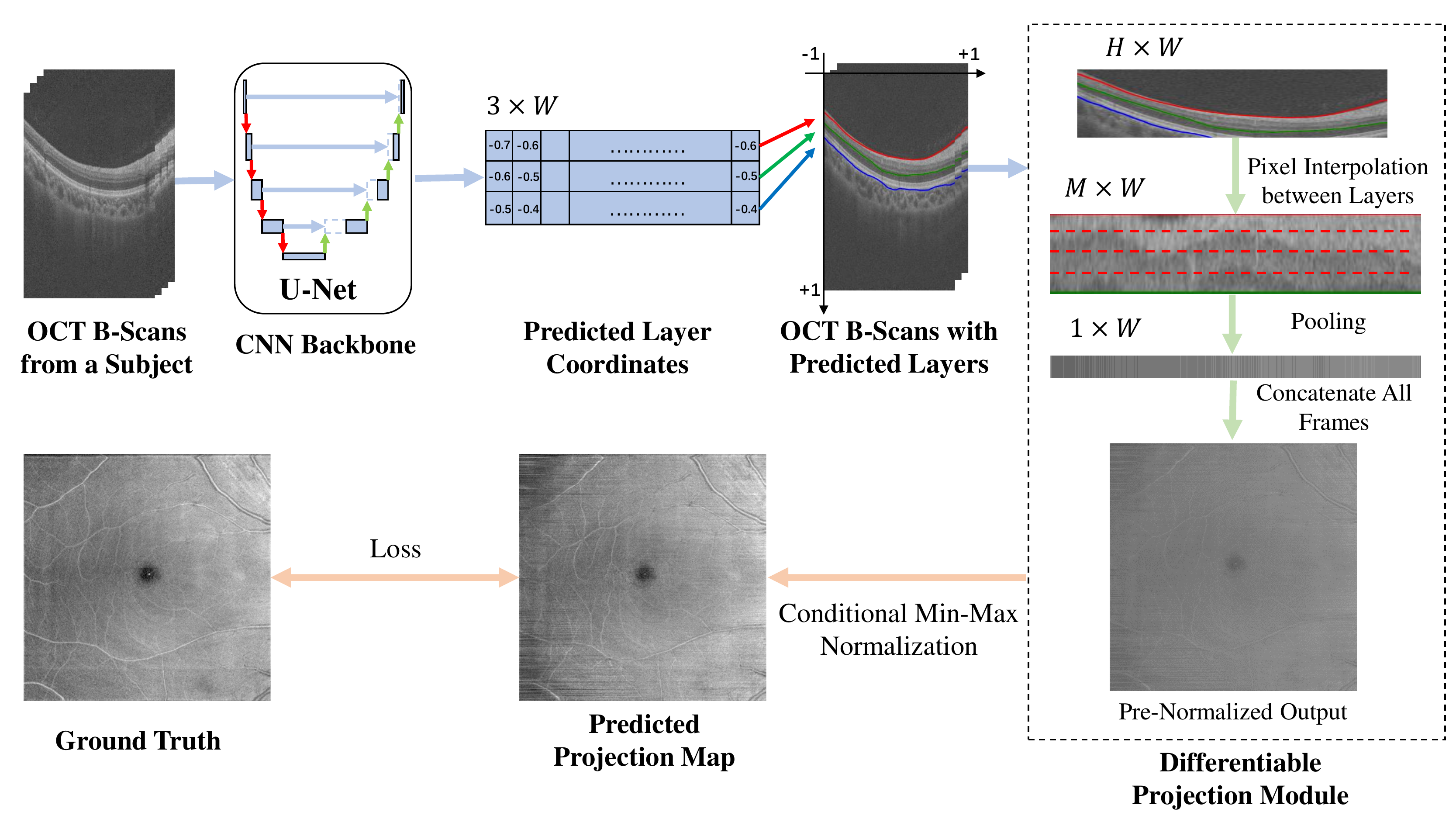}}

\end{minipage}

\caption{\textbf{Model Overview.} Our framework predicts projection maps (PMs) from OCT B-scans. Each slice of OCT B-scans produces a column in the PMs. The whole procedure is differentiable, thus enabling end-to-end training with the ground truth of PMs instead of retinal layer segmentation.}
\label{fig:overview}
\end{figure}

Optical coherence tomography (OCT) is an important modality in retina imaging thanks to its high resolution and non-invasiveness in 3D~\cite{huang1991optical}. 2D projection maps (PMs) between retinal layers from OCT B-scans are popular in retinal disease diagnosis, which provides information inlier retinal pathology invisible in the conventional fundus images~\cite{niemeijer2008vessel,chen2016high}. The 2D projection maps require retinal layer segmentation on each slice of B-scans, and then aggregate (\eg, average or max) pixels between certain layers. Traditional algorithms of retinal layer segmentation are typically based on prior knowledge of retinal layer structures, \eg, boundary tracking~\cite{lang2013retinal}, adaptive thresholding \cite{ishikawa2005macular}, gradient information in dual scales \cite{yang2010automated}, texture and shape analysis \cite{kajic2010robust}. However, hand-crafted algorithms are hard to generalize in the real world. Data-driven deep learning has been dominating medical image analysis~\cite{shen2017deep,litjens2017survey}. Researchers have developed deep learning-based methods for retinal layer segmentation~\cite{shah2018multiple,fang2017automatic} with proven superiority over traditional algorithms. Nevertheless, the performance of deep learning approaches is built upon numerous retinal layer segmentation labels, which could be especially tedious for hundreds of slices in OCT B-scans. 

In this study, we present an alternative strategy to obtain projection maps from OCT B-scans, WITHOUT explicit supervision of retinal layer segmentation. Instead, trained with pairs of OCT B-scans and the corresponding PMs, our end-to-end framework directly outputs the final target PMs. Although the PM ground truth is produced with retinal layer segmentation, paired B-scans and PMs could be easier to collect retrospectively as they are more likely to be stored in picture archiving and communication systems (PACS). As an example, the public OCTA-500 dataset~\cite{li2020ipn} used in this study (Fig.~\ref{fig:dataset} and Sec.~\ref{sec:dataset}) provides paired B-scans and PMs rather than retinal layer segmentation.

To our knowledge, this is the first study to predict projection maps from OCT B-scans in an end-to-end fashion. Unfortunately, a vanilla CNN without any geometric prior could only produce low-quality PMs. Therefore, we design a novel Differentiable Projection Module (DPM) to simulate the procedure of projection maps, inspired by spatial transformer networks \cite{jaderberg2015spatial}. As illustrated in Fig.~\ref{fig:overview}, instead of segmenting retinal layers explicitly, our CNN backbone predicts them implicitly as coordinates of curves in 2D views or surfaces in 3D, which are processed into areas/volumes by uniform point sampling between layers. They could be interpolated into pixels from source B-scans, and finally projected into PMs via (average/max) pooling. All the above operations are differentiable, which could be seamlessly integrated into neural networks and end-to-end trainable with the supervision of PM ground truth. The proposed method produces high-quality PMs with proven superiority over baselines (Sec.~\ref{sec:exp}). 
Independent from the clinical use cases, the introduced novel neural representation of areas/volumes between curves/surfaces, could be a technical contribution in shape modeling and geometric deep learning. 
To facilitate open research, our code is open source on GitHub\footnote{\url{https://github.com/dingdingdingyi/projection-from-OCT}}.

\section{Materials and Methods}
\label{sec:meth}

\subsection{Dataset Overview} \label{sec:dataset}
\begin{figure}

\begin{minipage}[b]{1.0\linewidth}
  \centering
  \centerline{\includegraphics[width=8.5cm]{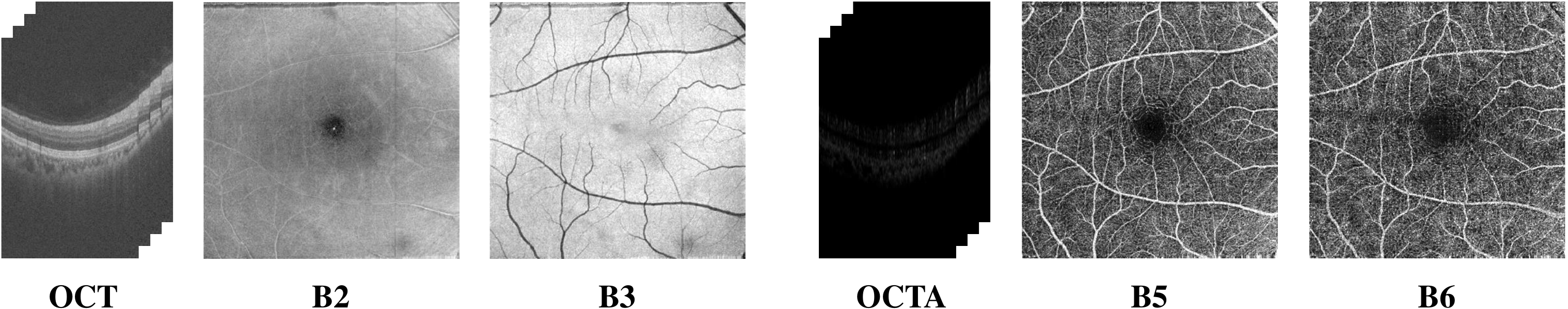}}

\end{minipage}

\caption{\textbf{The Public OCTA-500 Dataset~\cite{li2020ipn}.} Each sample contains the B-scans of OCT and OCTA together with their projection maps (B2, B3, B5 and B6). The retinal layer segmentation is NOT provided.}
\label{fig:dataset}
\end{figure}

We conducted experiments on the public OCTA-500 \cite{li2020ipn} dataset, specifically the OCTA\_6M subset, which contains 300 subjects with 6mm $\times$ 6mm FOV. As demonstrated in Fig.~\ref{fig:dataset}, each sample is paired OCT/OCTA B-scans and projection maps, where B2/B3 are projected from OCT and B5/B6 are from OCTA. Each slice of B-scans corresponds to a column in the projection maps. B2/B3 are the focus in this study. B2, projected by averaging pixels between ILM and OPL layers (\ie, the red and green curves in Fig.~\ref{fig:fulloct}), shows the vessels in the inner retina with high reflection; and B3, averagely projected between OPL and BM layers (\ie, the green and blue curves in Fig.~\ref{fig:fulloct}), shows the vessel shadows in the outer retina with low reflection. 
We use the official split, with 180/20/100 subjects (72,000/8,000/40,000 B-scans) for training, validation and evaluation, respectively.

\subsection{Method Overview}
\label{ssec:over}

As depicted in Fig.~\ref{fig:overview}, we first use a 2D U-Net \cite{ronneberger2015u} at a B-scan level to output coordinates of retinal layers, which are then processed by the proposed Differentiable Projection Module (DPM) to generate the projection maps. As the ground truth of the target PMs is min-max normalized over the whole PMs, we design a Conditional Min-Max (CMM) normalization trick during training to rescale the intensity values. The model parameters are trained via loss back-propagation between the prediction and ground truth of PMs.

\subsection{UNet Backbone}
Given an B-scan of $H\times W$, we use a downsampled image of $H_2\times W_2$ ($H_2=H/2, W_2=W/2$) as network input to reduce memory. The 2D U-Net backbone produces a feature map of $H_2\times W_2$. We apply a $H_2\times 1$ convolution with 3 output channels followed by a horizontal bilinear upsampling layer. The $3\times W$ output, representing predicted coordinates of the 3 retinal layers (\ie, ILM, OPL and BM) are mapped into $[-1,1]$ with $\mathit{tanh}$.

\subsection{Differentiable Projection Module (DPM)}

\begin{figure}

\begin{minipage}[b]{1.0\linewidth}
  \centering
  \centerline{\includegraphics[width=8.5cm]{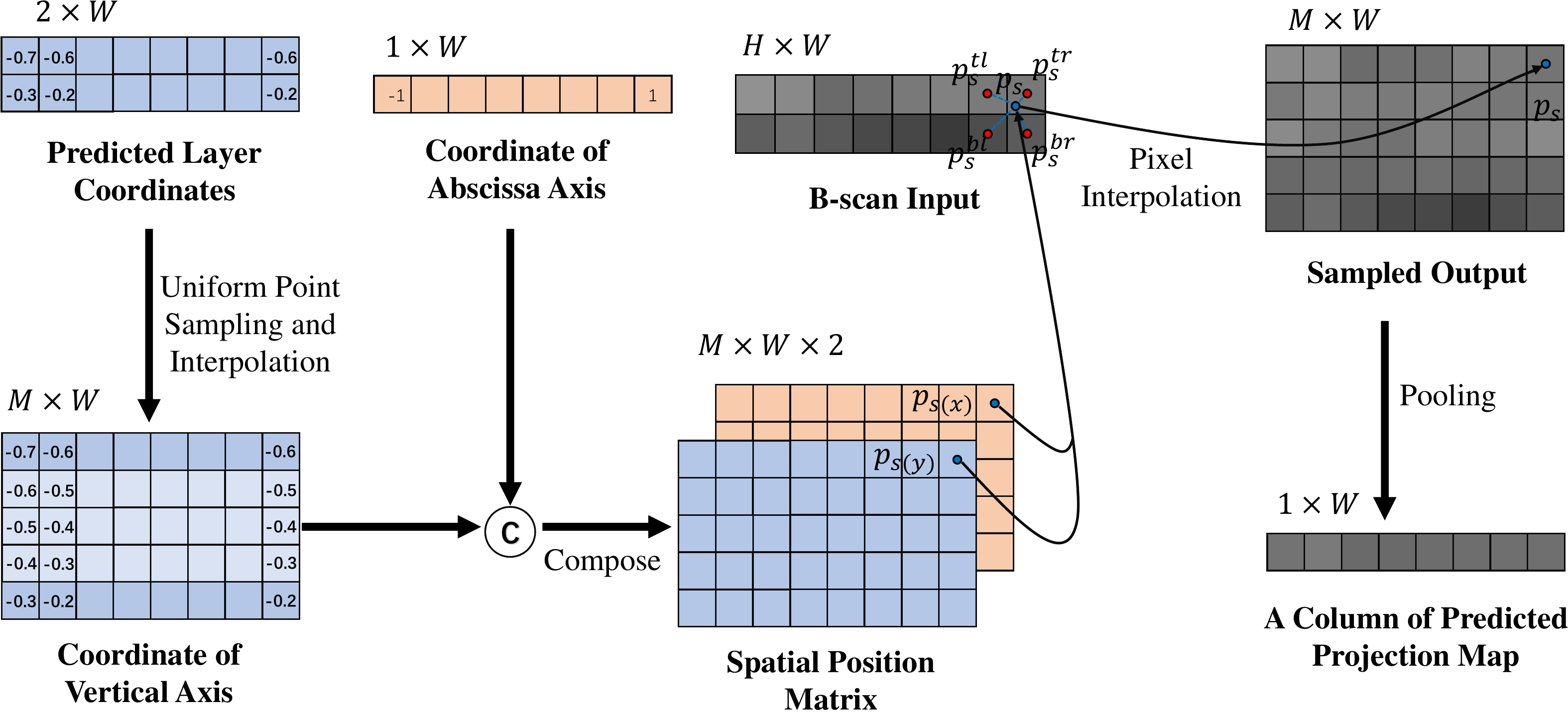}}

\end{minipage}

\caption{\textbf{Differentiable Projection Module.} Here, we illustrate 2 layers for simplicity. We uniformly sample points between the upper and lower predicted layer coordinates, whose output is denoted as a spatial position matrix of area/volume between retinal layers. The matrix could be interpolated into pixels from input B-scans, and finally (average/max) pooled into a column in projection maps. }

\label{fig:PM}
\end{figure}

For B2/B3, the projection maps are generated by averaging pixels between 2 layers (Sec.~\ref{sec:dataset}). For simplicity, we describe the Differentiable Projection Module (DPM) for 2 layers (\ie, $2\times W$). As shown in Fig.~\ref{fig:PM}, we uniformly sample $M$ points between the 2 layers (including the end points), which represent the vertical coordinates of sampled areas ($M\times W$). Then we assign a abscissa coordinate to each vertical coordinate, composing a spatial position matrix $G \in \mathbb{R}^{M \times W \times 2}$. For each coordinate, we could bilinearly interpolate the pixel value from the B-scan. This pixel interpolation operation introduced by the spatial transformer networks \cite{jaderberg2015spatial} could be easily implemented in modern deep learning frameworks, \eg, $\mathit{grid\_sample}$ in PyTorch~\cite{paszke2019pytorch}. The sampled output of $M \times W$ represents the warped image of pixels between the 2 layers. It could be averaged pooled vertically into $1\times W$, which is a column in the target projection map.

\subsection{Conditional Min-Max Normalization Trick (CMM)}
Our model produces a target PM column-by-column (\ie, slice-by-slice in B-scans); However, the PM ground truth provided in the OCTA-500 Dataset~\cite{li2020ipn} is min-max normalized over the whole PMs. The accurate min-max values of prediction could be only obtained given all slices, which is yet impossible during training due to memory constraint. Inspired by the \textit{auto-decoding} (NOT auto-encoding) training technique~\cite{park2019deepsdf}, we propose a Conditional Min-Max normalization trick to rescale the intensity of the projection maps:
\begin{equation}
\mathit{CMM}_i(I) = \frac{I - I_i^{min}}{I_i^{max} - I_i^{min}},
\end{equation}
where $I_i^{max}$ and $I_i^{min}$ are learnable parameters for the maximum and minimum values of the projected B-scans, conditional on the training subject ID $i$. Specifically, there are $180\times 2$ parameters for 180 training subjects in our study.

\subsection{Training and Inference}

The loss function of our model is composed of 2 terms: an L-1 loss $L_{1}$ and a feature loss $L_F$. The $L_{1}$ measures the pixel-wise similarity, while the $L_F$ measures structural similarity by computing the feature similarity using a pretrained network~\cite{johnson2016perceptual}. The project loss for B2 is defined as 
\begin{equation}
L_{B_2} = L_{1}(y_{i}, \mathit{CMM}_i(f(x_{i}))) + L_F(y_{i}, \mathit{CMM}_i(f(x_{i}))),
\end{equation}
where $y_{i}$ is the B2 ground truth, $f(x_{i})$ is the pre-normalzied prediction. When training the B$_{2}$ and B$_{3}$ projection maps simultaneously, the final loss is
\begin{equation}
\label{eq:1}
L = L_{B_{2}}+\lambda L_{B_{3}}.
\end{equation}

The $\mathit{CMM}$ normalization trick is only used during training. During inference, we predict pre-normalized outputs slice-by-slice, and then normalize it using the real min-max values on the predicted PMs.

\section{EXPERIMENTS}
\label{sec:exp}

\subsection{Experiment Settings and Details}
\label{ssec:set}

To validate the effectiveness of the proposed methods, we design the baselines: 1) CNN only: a vanilla CNN without DPM. 2) DPM only: an optimization-based DPM without the CNN part as a deep prior, \ie, optimizing the layer coordinates directly on each data pair. We use PSNR and SSIM to assess the predicted projection maps, where PSNR focuses on pixel similarity and SSIM focues on structural similarity.


Our framework is implemented in PyTorch~\cite{paszke2019pytorch}, all networks were trained on 4 NVIDIA RTX 2080 Ti with an Adam optimizer~\cite{kingma2014adam} using a batch size of 72. The learning rate starts with $1 \times 10^{-4}$ and exponentially decays with a ratio of $1 \times 10^{-2}$ after every epoch. We use $\lambda = 0.2$ in Eq.~\ref{eq:1} to balance the training for B$_{2}$ and B$_{3}$ with a simple grid search.

\begin{table}
	\renewcommand{\arraystretch}{1.3}
	\footnotesize
	\centering
	\begin{tabular}{lcccc}
    \toprule  
    Model& PSNR (B2) & SSIM (B2) & PSNR (B3) & SSIM (B3) \\
    \midrule  
    
    CNN only& 28.0424& 0.2869& 27.9896& 0.2332\\
    DPM only& 27.8129& 0.0663& 27.8800& 0.0683\\
    \midrule
    CNN + DPM& \textbf{28.7781}& \textbf{0.7575}& \textbf{28.7288}& \textbf{0.7195}\\
    w/o CMM& 28.2814& 0.5758& 28.3460& 0.6974\\ 
    \bottomrule
    \end{tabular}
	\caption{\textbf{Quantitative Results.} Evaluation of PSNR and SSIM results for the test set. DPM: Differentable Projection Module. CMM: Conditional Min-Max normalization trick. }
	\label{tab:table1}
\end{table}

\subsection{Quantitative Results}
\label{ssec:result}
We quantitatively compared the performance of our method against CNN-only method and DPM-only method to assess the effect of each component of our model, we use PSNR and SSIM for evaluation. Table \ref{tab:table1} gives a quantitative comparison of the results. From Table \ref{tab:table1}, we can observe that DPM-only method has the poorest performance, for this model has no component to understanding the input B-scan globally, just adjusts the layer segmentations blindly on the basis of the target projection map. CNN-only method also has a bad performance, we assume that this is because this method just uses a CNN network to predict the projection directly, ignoring the correct layered structure to get the target projection map.

In addition, we assessed the performance of CMM. The comparison demonstrates that adding CMM module is beneficial for learning layered structural information. However, the improvement of SSIM in B3 is not obvious, we assume that the reason is that B3 displays the vessel shadows in the outer retina with low reflection, in which vessel information is the dominance and the edges of tissues are distinct, so the accuracy of predicting ILM and BM layers does not have a great impact on projection B3 to a certain extent.

\begin{figure}[htb]

\begin{minipage}[b]{1.\linewidth}
  \centering
  \centerline{\includegraphics[width=8.5cm]{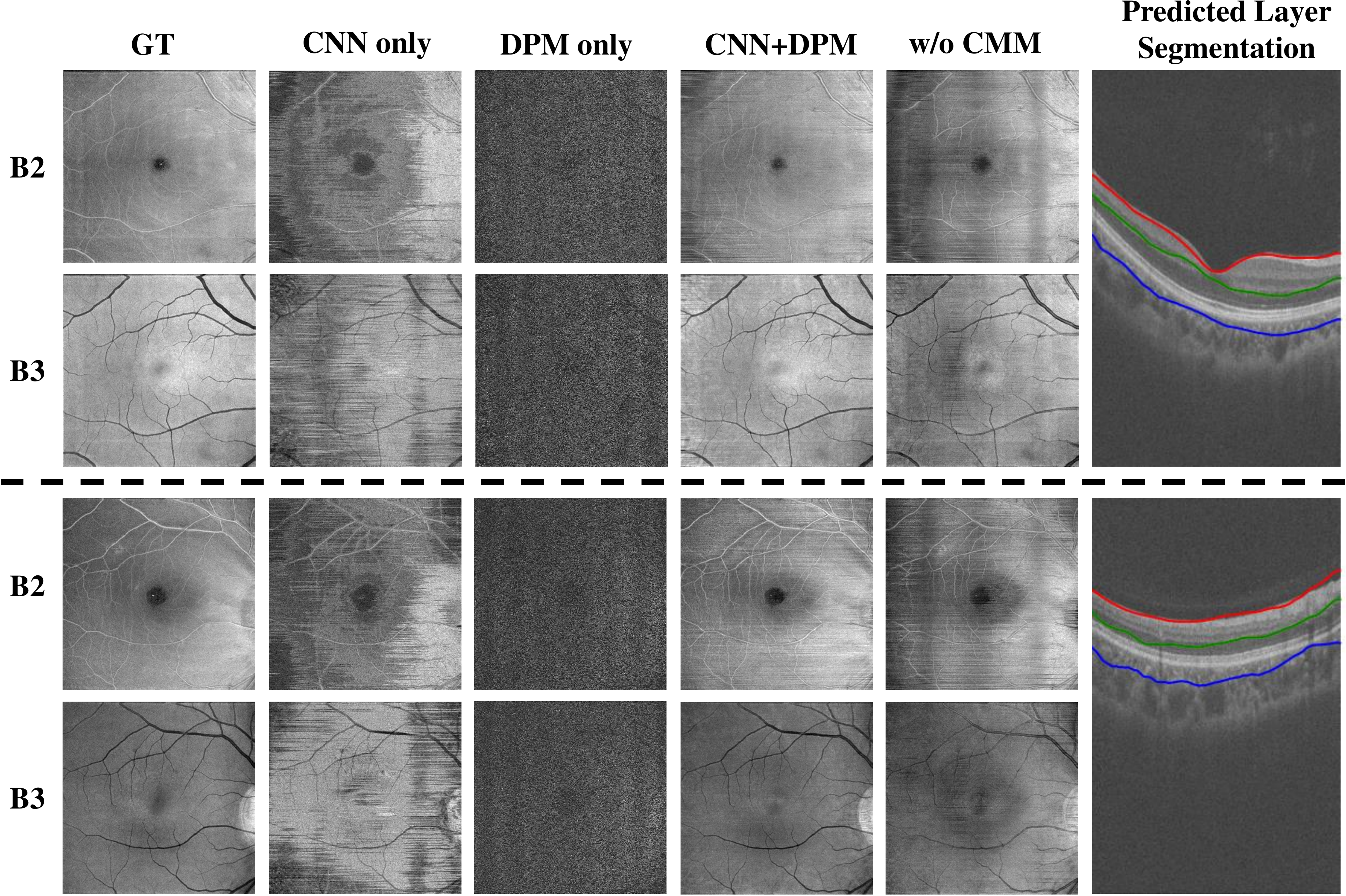}}

  \caption{\textbf{Qualitative Results.} 2 selected cases show qualitative comparison of different components. GT: Ground Truth. DPM: Differentable Projection Module. CMM: Conditional Min-Max. Best viewed on screen.}
  \label{fig:fulloct}
\end{minipage}
\end{figure}

\subsection{Qualitative Results}
In Fig. \ref{fig:fulloct}, we show projection maps obtained from CNN-only, DPM-only, our method (CNN+DPM) and without-CMM, and layers predicted by our method. The CNN-only model can produce most blood vessels successfully but has serious distortion in details. The DPM-only model totally fails to produce a projection map, the image is full of noise for the predicted layers have crossed, but the contours of some thick blood vessels are visible. From the last three images, we can clearly see that adding CMM module can help produce a clearer projection map and get more precise layers.

\subsection{Robustness on Extreme Cases}
\begin{figure}
\begin{minipage}[b]{.48\linewidth}
  \centering
  \centerline{\includegraphics[width=4.0cm]{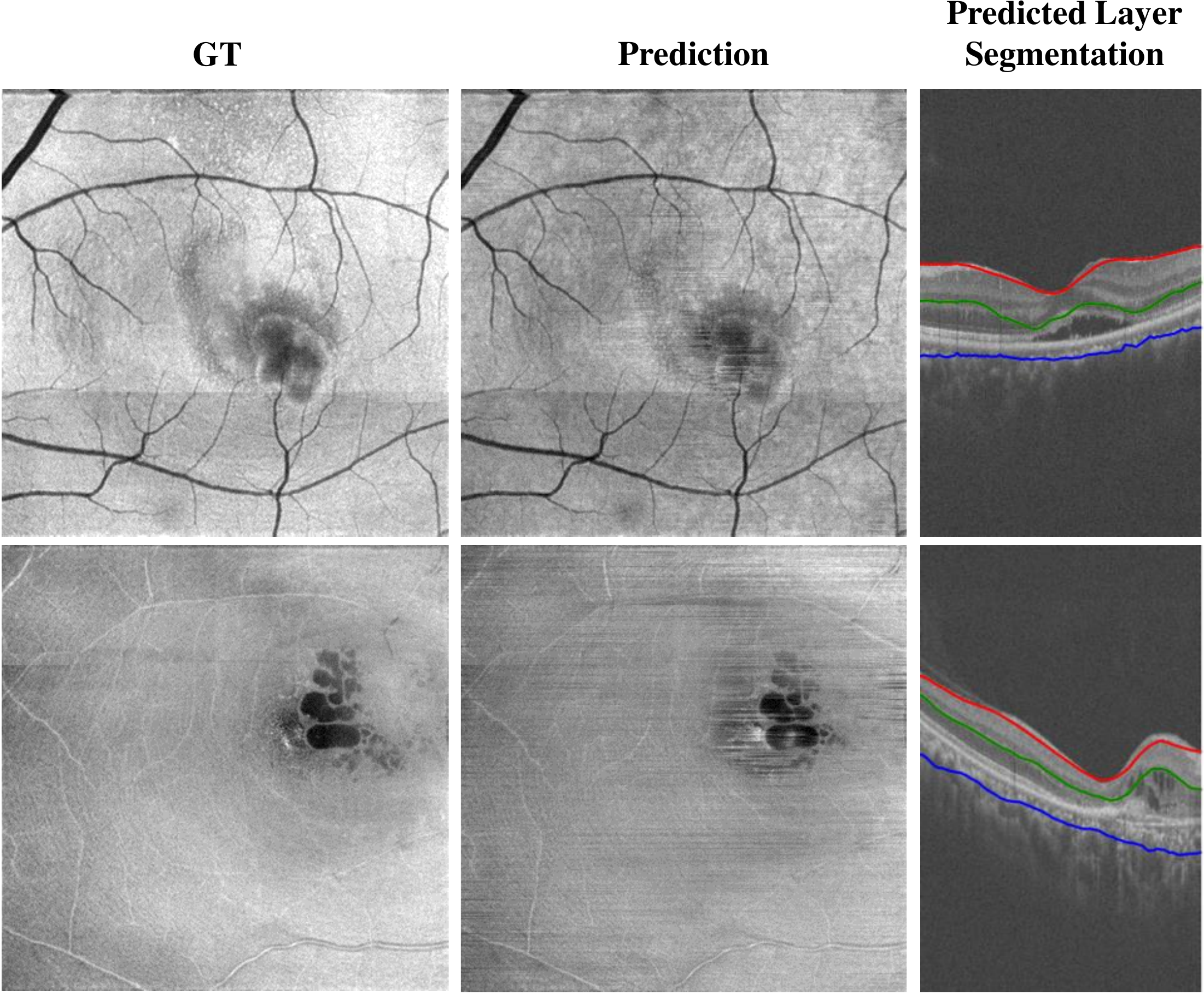}}
  \centerline{(a) Success on Extreme Cases}\medskip
\end{minipage}
\hfill
\begin{minipage}[b]{0.48\linewidth}
  \centering
  \centerline{\includegraphics[width=4.0cm]{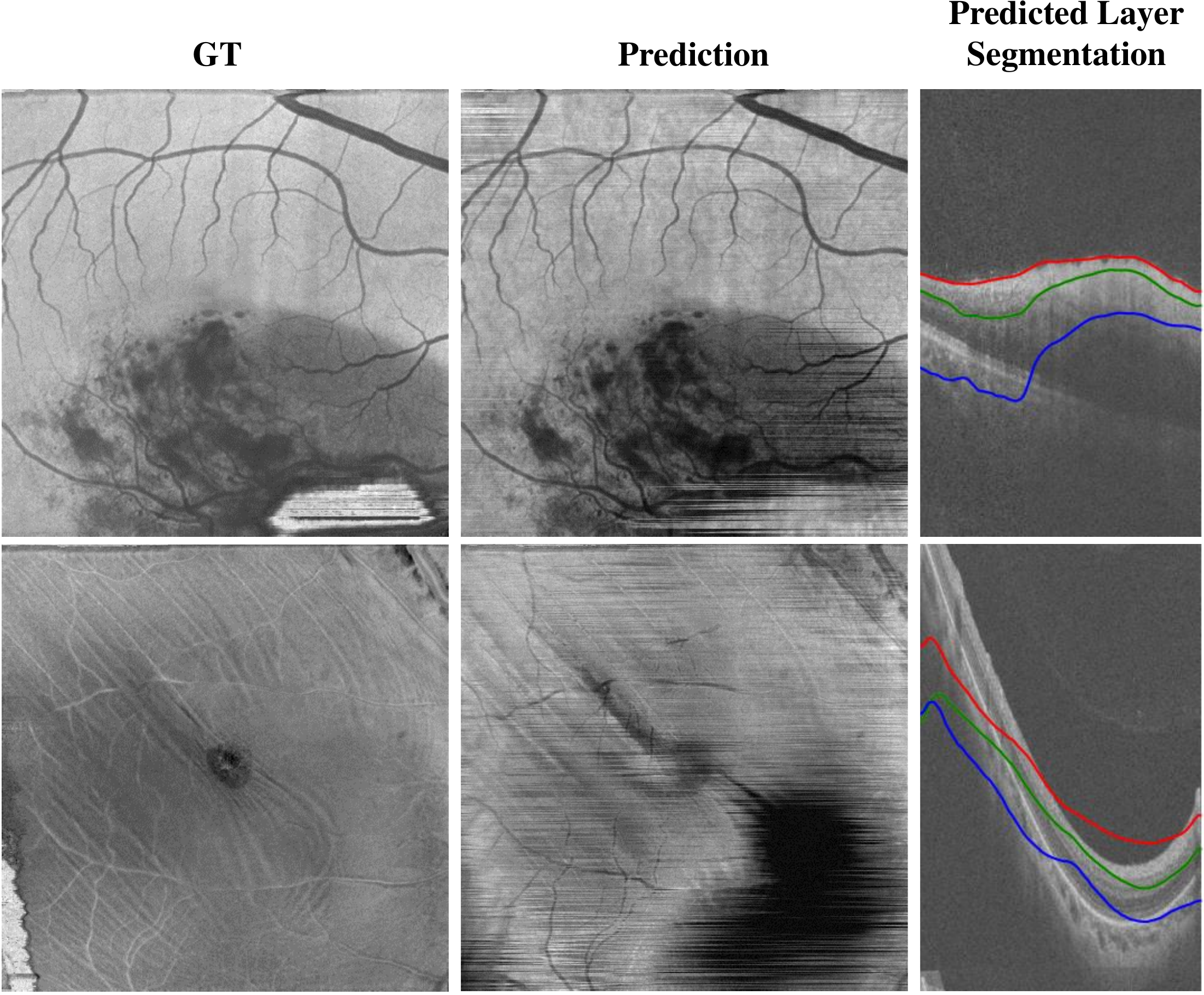}}
  \centerline{(b) Failure on Extreme Cases}\medskip
\end{minipage}
\caption{\textbf{Robustness on Extreme Cases.} (a) 2 cases with acceptable segmentation. (b) 2 cases with failure.}
\label{fig:daf}
\end{figure}

In Fig. \ref{fig:daf}~(a), we illustrate two cases with lesions, which are difficult to predict layers. In these cases, the boundaries of layers are blurred due to the presence of the diseased areas, which bring difficulties to the segmentation task. Our model tries to learn structural feature of diseased areas, and get acceptable results.

In Fig. \ref{fig:daf}~(b), we illustrate two typical failure cases. In the first example, our model fails to correctly segment the bottom layer of the B-scan, result in a serious distortion of the bottom part of the predicted projection map. The reason is that the bottom right part of the poor-quality original image lacks structural information. In the second example, the layers predicted by our model can not fit the target layers in B-scan very well, lead to generating a poor projection. The feature of the original B-scan is that the imaging of the retina is steep, which is rare in the used dataset, leading our model to learn more about flat B-scans and trying to predict smoother layers.

\subsection{Transferring from OCT to OCTA}
\begin{figure}
\begin{minipage}[b]{.48\linewidth}
  \centering
  \centerline{\includegraphics[width=4.0cm]{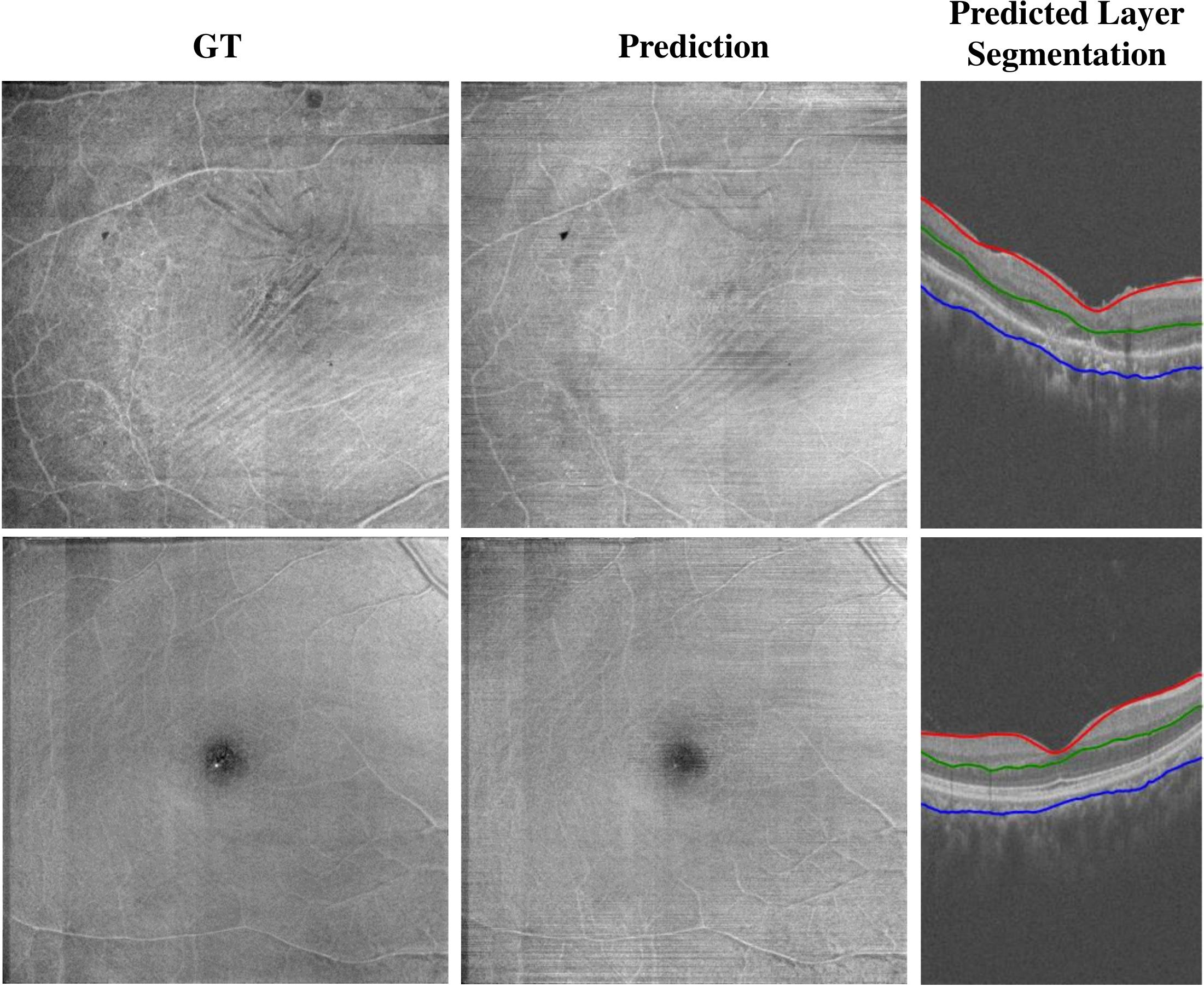}}
  \centerline{(a) Prediction on OCT}\medskip
\end{minipage}
\hfill
\begin{minipage}[b]{0.48\linewidth}
  \centering
  \centerline{\includegraphics[width=4.0cm]{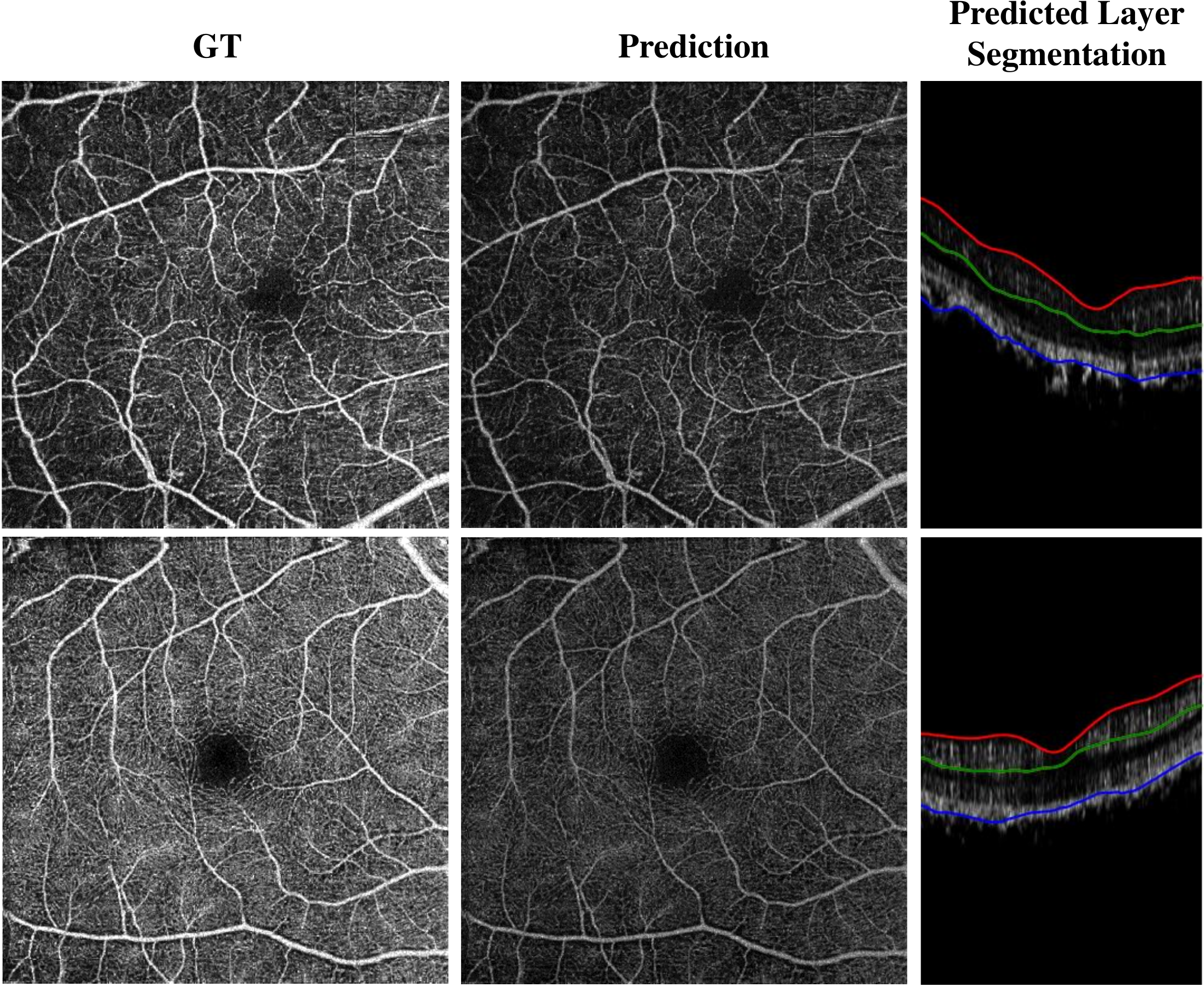}}
  \centerline{(b) Transfer to OCTA}\medskip
\end{minipage}
\caption{\textbf{Transferring from OCT to OCTA.} 2 cases of transferring the segmentation results from OCT to OCTA (upper and lower). (a) Predicted projections and segmentations of OCT. (b) Predicted projections and segmentations of OCTA in the same subject.}
\label{fig:octa}
\end{figure}

In this section, we try to transfer the segmentations to the OCTA B-scans. OCTA is  non-invasive technique that shows details of blood vessels that have low inherent contrast in OCT images. Given that OCTA images are generated from the OCT images, we directly transfer the layers of OCT to the OCTA, no need to train the model on OCTA again, and generate good quality projection maps. In Fig. \ref{fig:octa}, we illustrate two cases to show the results of transformation. Fig. \ref{fig:octa}~(b) shows the projection results of OCTA, we can see that the synthesized projection map generates the shape of blood vessels greatly and is structurally coherent with its corresponding ground-truth.

\section{Conclusion}
\label{sec:conclu}
In this study, we present a novel end-to-end framework to implicitly learn layer boundaries of an OCT B-scan from its projection map. Integrated with a 2D U-Net, the proposed end-to-end trainable CNN-DPM encourages the model to learn the structural feature of the input B-scan and predict layers from final projection maps. The qualitative and quantitative results demonstrate that the projection maps our framework generates have a similar texture with the real projection maps, which means that the layers are accurately predicted. However, for each B-scan of a subject, we predict layers independently, ignoring the structural continuous relationship between them, leading to streak distortion in generated projection. In our future work, we will improve our network to perform segmentation from the 3D OCT volume.

\section{COMPLIANCE WITH ETHICAL STANDARDS}
This research study was conducted retrospectively using human subject open-source data. Ethical approval was not required as confirmed by the license.

\section{ACKNOWLEDGEMENT}
This work was supported by National Science Foundation of China (U20B2072, 61976137). This work was also partially supported by Grant YG2021ZD18 from Shanghai Jiaotong University Medical Engineering Cross Research.
\bibliographystyle{IEEEbib}
\bibliography{strings,refs}

\end{document}